\documentstyle[a4,12pt,epsfig,graphicx]{article}
\begin{document}
\titlepage
\begin{flushright}
CERN-TH/2001-049\\
t01/024\\
hep-th/0102154 \\
\end{flushright}
\vskip 1cm
\begin{center}
{ \Large
\bf Non-BPS Instability in   Heterotic  M-theory}
\end{center}

\vskip 1cm
\begin{center}
{\large Ph. Brax\footnote{email: philippe.brax@cern.ch} }
\end{center}
\vskip 0.5cm
\begin{center}
Theoretical Physics Division, CERN\\
CH-1211 Geneva 23\footnote { On leave of absence from  Service de Physique Th\'eorique, 
CEA-Saclay F-91191 Gif/Yvette Cedex, France}\\
\end{center}

\vskip 2cm
\begin{center}
{\large \bf Abstract}
\end{center}
\vskip .3in \baselineskip10pt{We 
study the warped geometry of heterotic M-Theory in five dimensions
where five-branes are included in the bulk. 
Five-branes wrapping holomorphic curves lead to BPS configurations 
where the junction conditions are automatically satisfied. We consider
five-branes wrapped around non-supersymmetric cycles and show that
the configuration is unstable. We describe explicitly the resulting time-dependent
geometry where the bulk five-branes move towards the Horova-Witten boundary walls.
The five-branes collide with the boundary walls in a finite time resulting
in the   restoration of supersymmetry.
}
\bigskip
\vskip 1 cm

\noindent
\newpage
\baselineskip=1.5\baselineskip
\section{Introduction}
Five dimensional models with warped geometries have played a significant role 
in high energy physics and cosmology during the past few years.
Many puzzles in four dimensions can  been tackled by extending space-time to five dimensions\cite{kac}.
In most settings the fifth dimension is either an interval in models
inspired by the Horava-Witten theory\cite{HW} or a half-line with an AdS geometry for the Randall-Sundrum
proposal\cite {RS}. The phenomenological applications are numerous
ranging  from the hierarchy problem\cite{RSI}  to the cosmological constant problem\cite{Dim}.

Another guise of five dimensional models has been motivated by the AdS/CFT  correspondence( for a review see  \cite{ya}).  The search for a supersymmetric
version of the RS scenario  has led to the concept of supergravity in singular spaces\cite{ka,pok}.
A particularly interesting class of supergravity models coupled to boundaries emerges
from the compactification of M-theory on Calabi-Yau three-folds\cite {luk}. In five dimensions this leads to
gauged supergravity theories with  background fluxes. The origin of the background fluxes springs from the
modification of the Bianchi identities due to the Horava-Witten boundaries and the inclusion of
five-branes in the bulk\cite{ov}. The fluxes lead to the existence of potential terms for the vector multiplets.

The five dimensional solutions of the supergravity equations of motion have been
widely studied (for a review see \cite{ovr}). In particular one  finds BPS configurations preserving $N=1$ supersymmetry in four
dimensions. It also conspicuous to find cosmological solutions with a time dependent background \cite{cosm}.
These solutions should have some relevance to the physics of the early universe.

Recently non-BPS configurations have been extensively studied both from a 
theoretical and phenomenological point of view\cite{senreview,sentach} and might eventually lead to 
a better understanding of the origin
of supersymmetry breaking. One of the purpose of the present letter is
to consider non-BPS configurations in the context of the strongly coupled 
heterotic string theory\cite{sol}.
In particular we shall be concerned with unstable  configurations 
resulting from the presence of five-branes wrapping non-supersymmetric cycles in the bulk.

In a first section we recall the necessary ingredients of five 
dimensional supergravity and its link to heterotic M-theory.
We describe the BPS
situation with  five-branes wrapping supersymmetric cycles. We pay particular attention
to the boundary conditions and show that the BPS property guarantees that the junction conditions are automatically satisfied. We then break supersymmetry by wrapping five-branes around non-supersymmetric cycles.
After recalling the topological features of such configurations, we show that the modified boundary
action on the five-branes leads to an instability. We find explicit solutions of the equations of motion
where the five-branes move towards the boundary walls. The collision occurs within a time
proportional to the fifth-dimension radius.

\section{Bulk Five-Branes and BPS Configurations}
We compactify  M-theory\cite{HW} on a Calabi-Yau three-fold with background fluxes  switched on\cite{luk}.
The compactification of 
eleven  dimensional supergravity on a Calabi-Yau threefold with Hodge numbers $(h^{(1,1)},h^{(2,1)})$ leads to an 
N=2 D=5 supergravity theory with $h^{(1,1)}-1$ vector multiplets and $h^{(2,1)}+1$ hypermultiplets.
In particular there is always one universal hypermultiplet containing the volume of the
Calabi-Yau manifold $Y$. The other hypermultiplets belonging to a quaternionic
moduli space will not play a role here. Switching on background fluxes leads to a potential for the vector multiplets. The resulting supergravity is gauged
with the axion field charged under a $U(1)_R$ symmetry. One can introduce  five-branes
spread along four of the five non-compact directions and two of the internal dimensions.
For five-branes wrapping holomorphic curves  
the resulting configuration is BPS and breaks half of the supersymmetries. 

Let us now be a bit more specific. The compactifying manifold $Y$ lives along the
$x_A,\ A=5\dots 10,$ coordinates while space time is along $x_i, \ i=0\dots 3,$ and $x_5$.
The $x_5$-axis is  a Z2 orbifold under $x_5\to -x_5$ that we identify with
the interval $[0,+\pi\rho]$. 
We also include  $N$ five-branes $M_5^i$
extended along the four non-compact dimensions
$x_i$ and wrapped around holomorphic curves  $\Sigma^i$ in $Y$. The Bianchi 
identity  for the four form $G_4$ of eleven  dimensional 
supergravity is modified due to the presence of the five-brane 
\begin{equation}
dG_4=4\sqrt 2 (\frac{\kappa}{4\pi})^{2/3}[\delta_{M_5}-\delta_{0}dx_5\wedge J^{(1)}-\delta_{\pi \rho}dx_5\wedge J^{(2)}]
\end{equation}
with 
\begin{equation}
J^{(i)}=\frac{1}{16\pi^2}(
\hbox{Tr}(F^{(i)}\wedge F^{(i)})-\frac{1}{2}\hbox{Tr}(R\wedge R)).
\end{equation}
Due to supersymmetry preservation each of the Horava-Witten planes carries
a holomorphic vector bundle characterized by the two forms $F^{(i)},\ i=1,2.$ 
We will choose $F^{(2)}=0$ and consider the first plane at $x_5=0$ as our brane-world. 
The fundamental class $\delta_{M_5}$ is defined by 
\begin{equation}
\int_{M_5}f=\int_{M_{11}}f\wedge \delta_{M_5}
\end{equation}
and  can be written as
\begin{equation}
\delta_{M_5}=\sum_{i=1}^N\delta_{x^i_0} dx_5\wedge \delta_{\Sigma^i}
\end{equation}
where $x^i_0$ are  the coordinates of the five-branes. In appropriate
units $\delta_{\Sigma^i}$ has  dimension three in eleven dimensional Planck units. 

Due to the compactness of the $x_5$ direction there is a topological condition to be satisfied
\begin{equation}
\delta_{\Sigma}\equiv \frac{1}{16\pi^2}(\hbox{Tr}(F^{(1)}\wedge F^{(1)})-\hbox{Tr}(R\wedge R))
\label{top}
\end{equation}
as cohomology classes where $\delta_{\Sigma}=\sum_{i=1}^N\delta_{\Sigma^i}$. 
This explicitly determines the homology class of the two dimensional  surface
$\Sigma$.

An explicit solution to the Bianchi identity is then
\begin{equation}
G_4= 2\sqrt 2(\frac{\kappa}{4\pi})^{2/3} \sum_{i=0}^N \epsilon_{ x_0^i}\ \delta_{\Sigma^i}
\end{equation}
where $\epsilon_{x_0^i}$ jumps from -1 to 1 at $x_0^i$  and we have  used  the notation
\begin{equation}
\delta_{\Sigma^0}=-\frac{1}{16\pi^2}(
\hbox{Tr}(F^{(1)}\wedge F^{(1)})-\frac{1}{2}\hbox{Tr}(R\wedge R)).
\end{equation}
In each interval separating the $i$-th and the $(i+1)$-th
five-branes  there are  background  magnetic charges defined by
\begin{equation}
\alpha_I^i=\frac{ \sqrt 2 \epsilon_S}{\rho}\int_{C_I}\sum_{j=0}^{i}\delta_{\Sigma_j}
\end{equation}
where the four-manifolds $C_I$  are Poincar\'e duals to the 
the $\omega_I$'s forming a basis of the $h^{(1,1)}$ holomorphic two-forms.
We have introduced the expansion parameter $\epsilon_S=(\frac{\kappa}{4\pi})^{2/3}\frac{2\pi \rho}{v^{2/3}}$
and $v$ is  the volume of $Y$.
The effective action obtained by substituting in the 
eleven dimensional supergravity action depends
crucially on these magnetic charges.

The vector multiplets  follow from the 
expansion of the Kahler form $\omega $
\begin{equation}
\omega=t^I\omega_I.
\end{equation}
The  volume modulus of the Calabi-Yau manifold is given by
\begin{equation}
{\cal V}=\frac{1}{6}\int_{Y}\omega\wedge\omega\wedge\omega.
\end{equation}
Defining 
\begin{equation}
t^I={\cal V}^{1/3}X^I
\end{equation}
the scalars in the vector multiplets parameterize the solutions
of
\begin{equation}
C_{IJK}X^IX^JX^K=6
\label{con}
\end{equation}
where $C_{IJK}$ are the intersection numbers $\int_{Y}\omega_I
\wedge\omega_J\wedge\omega_K$.
The volume modulus belongs to the universal hypermultiplet. 
The low energy bosonic action takes the form of a non-linear sigma
model with the metric defined by
\begin{equation}
G_{IJ}=-\frac{1}{2}C_{IJK}X^K+\frac{1}{8}(C_{ILM}X^LX^M)(C_{JPQ}X^PX^Q).
\end{equation}
In the Einstein frame $ds^2_E={\cal V}^{2/3}ds^2_{str}$ the action for the $i$-th interval reads\cite{luk}
\begin{equation}
S_{bulk}^i=-\frac{1}{2\kappa_5^2}\int d^5x \sqrt{-g_E^{(5)}}(R+G_{IJ}\partial_{\mu}X^I\partial^{\mu}X^J
-\frac{1}{2{\cal V}^2}(\partial {\cal V})^2 -\frac{1}{2{\cal V}^2}\alpha_I^i\alpha_J^iG^{IJ}(X)).
\end{equation}
The last term is the potential for the scalars in the vector multiplets.

Let us now consider the boundary actions. The  boundary wall action reads \cite{luk} 
\begin{equation}
S_{B}^0=\frac{\sqrt 2}{\kappa_5^2}\int d^4x \sqrt{-g_E^{(4)}}\frac{\alpha_I^0 X^I}{\cal V}
\end{equation}
in the Einstein frame.
The action on the three branes resulting from the bulk five-branes depends on the nature of the
surface $\Sigma$.
For BPS configurations  the BPS bound for five-branes is saturated implying the equality
between the central  charge and the five-brane tension. 
This leads to the action for each five-brane 
\begin{equation}
S_B^i=T_5 \int \sqrt{-g^{(4)}_{str}}\hbox{Vol}(\Sigma^i)
\end{equation}
in the string frame. Now using\cite{green} 
\begin{equation}
T_5=\frac{2\pi}{({4\pi \kappa})^{2/3}}
\end{equation}
and 
\begin{equation}
\hbox{Vol}(\Sigma^i)=v^{1/3}\int_{\Sigma^i}\omega
\end{equation}
one obtains the boundary action
\begin{equation}
S_{B}^i=\frac{1}{\sqrt 2 \kappa_5^2}\int d^4x \sqrt{-g_E^{(4)}}\frac{[\alpha_I^i] X^I}{\cal V}
\end{equation} 
with $[\alpha_I^i]=\alpha_I^i-\alpha_I^{(i-1)}$. Notice that this boundary action has the same functional form as
the boundary wall action up to a factor of two. In the following section we will concentrate
on the solutions of the equations of motions with a particular emphasis on the junction conditions.

\section{Non-BPS Configurations}

Before describing the non-BPS configurations we  analyse the solutions of the equations of motions
in the BPS case. 
Let us concentrate on the following warped geometry
\begin{equation}
ds^2=e^{2A(x_5)}dx_{//}^2 + dx_5^2
\end{equation}
and consider $x_5$-dependent fields only. From the action one can read off the junction conditions at 
the origin
\begin{eqnarray}
&&[\frac{d{\cal V}}{dx_5} ]_0={2\sqrt 2 (\alpha^0. X)}\vert_0\nonumber  \\
&&[\frac{d A}{dx_5} ]_0=\frac{\sqrt2}{3}\frac{(\alpha^0.X)}{{\cal V}}\vert_0\nonumber  \\
&&[\frac{dX_I}{dx_5} ]_0=\frac{\sqrt 2}{\cal V}(\alpha^0_I-\frac{2}{3}(\alpha^0.X)X_I)\vert_0\nonumber  \\
\end{eqnarray}
where in the last equation we have used a Lagrange multiplier to impose the constraint (\ref{con}).
Similarly the junction conditions at the one of the bulk five-branes read
\begin{eqnarray}
&&[\frac{d{\cal V}}{dx_5}]_{x^i_0}=\sqrt 2 ([\alpha^i].X)\vert_{x_0^i}\nonumber \\
&&[\frac{dA}{dx_5}]_{x_0^i}=\frac{1}{3\sqrt2} \frac{([\alpha^i].X)}{{\cal V}}\vert_{x_0^i}\nonumber \\
&&[\frac{dX_I}{dx_5}]_{x_0^i}=\frac{1}{\sqrt 2{\cal V}}([\alpha^i_I]-\frac{2}{3}([\alpha^i].X)X_I)\vert_{x_0^i}\nonumber \\
\end{eqnarray}
where the latter differ from the former by a factor of two.
Due to the $Z2$ action one has $[f]_0=2f\vert_0$ relating the jump at the origin to twice the
value at the origin.

One of the features of BPS configurations is that the junctions conditions are automatically satisfied.
This can be seen from the BPS equations deduced from the fermionic supersymmetry variations
\begin{eqnarray}
&&\frac{d{\cal V}}{dx_5}=\sqrt 2 (\alpha^i.X)\nonumber \\
&&\frac{dA}{dx_5}=\frac{1}{3\sqrt 2} \frac{(\alpha^i.X)}{ {\cal V}}\nonumber \\
&&\frac{dX_I}{dx_5}=\frac{1}{\sqrt 2{\cal V}}(\alpha^i_I-\frac{2}{3}(\alpha^i.X)X_I)\nonumber \\
\end{eqnarray}
in each interval. 
Combining these equations one gets\cite{luk}
\begin{equation}
\frac{d({\cal V}^{1/3}X_I)}{dx_5}={\cal V}^{-2/3}\frac{\alpha_I^i}{\sqrt 2}
\end{equation}
from which we deduce that in each interval
\begin{equation}
C_{IJK}t^Jt^K=2\sqrt 2 {\cal V}^{1/3} \alpha_I^i y + C_I
\label{quad}
\end{equation}
where we have introduced $dy={\cal V}^{-2/3}dx_5$. Inverting (\ref {quad}) for $t^I$ 
gives  
\begin{equation}
{\cal V}=\frac{1}{6}C_{IJK}t^It^J t^K
\end{equation}
and the metric
\begin{equation}
ds^2={\cal V}^{1/3}dx_{//}^2 +{\cal V}^{4/3}dy^2.
\end{equation}
The position of the five-branes is not constrained reflecting the
no-force condition.

The non-BPS configurations appear when the cohomology class $\delta_{\Sigma}$ given by (\ref{top}) is not effective, i.e.
when the expansion 
\begin{equation}
\delta_{\Sigma}=\sum_{i=1}^M a_i \delta_{C^i},
\end{equation}
in terms of the classes $\delta_{C^i}$  of  holomorphic curves $C^i$,
contains both positive and negative integers. This implies that there is no
holomorphic curve whose class coincides with $\delta_{\Sigma}$.  This topological characterization
of non-supersymmetric cycle has several physical consequences. On the one hand the no-force condition
between BPS five-branes is no longer valid. Writing
\begin{equation}
\delta_{\Sigma}=[A]-[B]
\end{equation}
where $[A]$ and $[B]$ are effective cycles, i.e. there exist  holomorphic curves $A$ (resp. $B$ )
whose classes  are  $[A]$ (resp. $[B]$),
one expects  that separating the five-branes wrapped around $A$
from the anti-five-branes wrapped around $B$ is  energetically disfavoured. 
Therefore we will consider that the five-branes coincide and wrap a single surface in 
the class $\delta_{\Sigma}$.
Within the homology class dual to $\delta_{\Sigma}$ we consider the surface $S$ whose volume is minimum. 
As the brane tension is  minimal this  configuration is stable for a given Calabi-Yau manifold $Y$, i.e.
the five brane wrapped around $S$ gives rise to a stable non-BPS brane. 
The BPS bound states that the tension of the five brane wrapped around $S$  is  bounded from below
by the central charge 
\begin{equation}
\vert Q\vert =T_5v^{1/3}\int d^4 x \sqrt  {-g_{str}^{(4)}}\ \ \vert \int_{\Sigma}\omega\vert. 
\end{equation}
Equality would imply that $S$ is a calibrated surface\cite{beck,gib}, i.e. a holomorphic curve, realizing a BPS configuration. 
Such stable non-BPS branes are sensitive to deformations of the Calabi-Yau manifold $Y$\cite{senK}. 

Let us denote by $T>1$ the ratio beween the tension and the central charge. This leads to the boundary action
\begin{equation}
S_{B}^i=\frac{1}{\sqrt 2 \kappa_5^2}\int d^4x \sqrt{-g_E^{(4)}}\frac{[\alpha_I] X^I}{\cal V}T
\end{equation} 
with
\begin{equation}
\alpha_I=\frac{ \sqrt 2 \epsilon_S}{\rho}\int_{C_I}\delta_{\Sigma}.
\end{equation}
The effect of supersymmetry breaking is to modify the boundary action.
In particular the bulk equations of motion are still the same as before supersymmetry breaking.
Therefore the solutions of the bulk equations of motion are not modified  by the
supersymmetry breaking mechanism. 
We first assume that $T$ is independent of ${\cal V}$ and $X^I$.
The only effect of supersymmetry breaking is to
alter  the boundary conditions
\begin{eqnarray}
&&[{\partial_n{\cal V}}]_{x_0}=\sqrt 2 T([\alpha].X)\vert_{x_0}\nonumber \\
&&[{\partial_n A}]_{x_0}= \frac{T}{3\sqrt 2}\frac{([\alpha].X)}{ {\cal V}}\vert_{x_0}\nonumber \\
&&[{\partial_n X_I}]_{x_0}=\frac{T}{\sqrt 2 {\cal V}}([\alpha_I]-\frac{2}{3}([\alpha].X)X_I)\vert_{x_0}\nonumber \\
\end{eqnarray}
where $\partial_n$ is the normal derivative. 
The boundary conditions deduced from the bulk solutions
do not match with the boundary conditions arising from the non-BPS brane action.
Such a discrepancy has already been analysed in the context of supergravity in singular spaces \cite{brax}.
In particular we expect that the presence of a non-BPS brane  destabilizes the
vacuum. The perfect balance between the gravitational and 
scalar forces  disappears and the non-BPS brane moves towards the boundary walls.

We can  generate time-dependent conformally flat solutions from the static
solutions\cite{brax, pierre}. This is 
most easily achieved by using a boost along
the $x_5$ direction.  To do so we  first introduce conformal coordinates
so that the metric becomes
\begin{equation}
ds^2_5=a^2(u)(dx^2+du^2).
\end{equation}
where
\begin{equation}
a^2={\cal V}^{1/3}, \ 
du={\cal V}^{1/2} dy.
\end{equation}
Under a boost and a rescaling the new solutions of the bulk equations of motions are 
\begin{eqnarray}
\tilde A(u,\eta)&=&  A(u+h\eta,\frac{\alpha_I^i}{\sqrt{1-h^2}})\nonumber \\
\tilde{\cal V} (u,\eta)&=& {\cal V}(u+h\eta,\frac{\alpha_I^i}{\sqrt{1-h^2}} )\nonumber \\
\tilde X_I(u,\eta)&=& X_I(u+h\eta,\frac{\alpha_I^i}{\sqrt{1-h^2}} )\nonumber \\
\label{sol}
\end{eqnarray}
where $ \eta$ is the conformal time. We have displayed the explicit dependence on the magnetic charges $\alpha_I^i$.
One can  now use the BPS equations satisfied by $(A,{\cal V}, X_I)$ to deduce that
\begin{eqnarray}
&&{\partial_n\tilde {\cal V}}=\frac{\sqrt 2}{\sqrt {1-h^2}} (\alpha^i.\tilde X)
\nonumber \\
&&{\partial_n \tilde A}=\frac{1}{3 \sqrt 2\sqrt {1-h^2}} \frac{(\alpha^i.\tilde X)}{ \tilde {\cal V}}\nonumber \\
&&{\partial_n \tilde X_I}=\frac{1}{\sqrt 2 \sqrt {1-h^2}}\ \frac{1}{\tilde {\cal V}}
\ (\alpha^i_I-\frac{2}{3}(\alpha^i.\tilde X)\tilde X_I)\nonumber \\
\label{T}
\end{eqnarray}
in each of the two intervals.
Evaluating the jumps at $u_0$, the fixed coordinate of the non-BPS brane, and  using (\ref {T}) one finds 
that the boundary conditions are automatically satisfied 
provided that
\begin{equation}
h=\pm \frac{\sqrt{T^2-1}}{T}.
\end{equation}
Notice that this requires $T\ge 1$. This is exactly the BPS bound with
a static solution only in the BPS case.

After applying the boost the two boundary walls are moving and the non-BPS brane is static. 
By reverting to the original coordinates one finds that the solution  describes a moving non-BPS brane
surrounded by two static boundary walls. As the bulk equations of motions are not modified by the presence
of the non-BPS brane the boundary conditions at the two boundary walls are automatically satisfied.
The moving non-BPS brane eventually hits the boundary walls in a finite time determined by the speed $h$.
For a generic supersymmetry breaking parameter the life-time of the non-BPS brane is of the order of the size of the 
fifth dimension. 

When the supersymmetry breaking $T$ depends on ${\cal V}$ and $X^I$, the bulk equations are still
satisfied but the boundary conditions cannot be simply fulfilled by applying a boost in the fifth
direction. In \cite{sol} the case where supersymmetry is broken in this fashion
by one of the boundary walls was  considered. It has been shown that static solutions 
would require a fine-tuning of the radius of the extra dimension and of the magnetic charges.
The latter being unlikely because this requires tuning continuous and discrete variables.
From our point of view the reason for the non-existence of static solution follows from the absence
of balance between the forces on the non-BPS brane. It seems likely that an appropriate
change of variables performed on the bulk solutions will imply  the matching of the boundary
conditions in this more general setting. The time dependence of the resulting solution would describe  the
motion of the non-BPS brane towards the boundary walls.

Let us now briefly discuss the fate of the non-BPS brane after hitting one 
of the boundary walls. First of all the case of BPS five branes merging with the boundary walls
has been extensively studied\cite{wit}.
In particular one finds  small instanton transitions where the gauge bundle is modified.
In the non-BPS case the flux conservation condition indicates that supersymmetry configurations
such as the standard embedding case can  spontaneously appear after the collision. This leads 
to a restoration of supersymmetry.

\section{Conclusion}

We have described the sharp difference between BPS and non-BPS 
configurations in heterotic M-Theory. In particular we have explicitly shown that
the absence of balance between the gravitational and scalar forces leads to the motion
of the non-BPS five-branes towards the boundary walls. 
Due to the finite size of the fifth dimension  the life-time of the non-BPS
configuration is measured in units of the eleven dimensional Planck length.
Nevertheless one may use the resulting configuration in a phenomenological
way by considering that the size of the extra dimension is large. Indeed the BPS
condition on the boundary walls guarantees the absence of a radion potential and therefore
allows to consider an arbitrarily large extra dimension. Moreover by tuning the supersymmetry
breaking scale one may  consider the speed of the extra dimension to be sufficiently small
to allow for an adiabatic treatment. The resulting scenario might  be useful
in order to  study the  supersymmetry breaking induced by non-BPS branes in brane-world
models. 

\section{Acknowledgements}
I would like to thank C. Grojean and D. Waldram for many suggestions and useful comments.


\newpage

\begin{thebibliography}{999}

\bibitem{kac} S. Kachru `` Lectures on Warped Compactifications and Stringy Brane Constructions'',
hep-th/0009247.

\bibitem{HW} P. Horava and E. Witten, {\it Nucl. Phys} {\bf B460} (1996) 506, P. Horava and E. Witten, {\it Nucl. Phys} {\bf B475} (1996) 94.


\bibitem{RS} L. Randall and R. Sundrum, {\it Phys. Rev. Lett.} {\bf 83} (1999) 4690.

\bibitem{RSI} L. Randall and R. Sundrum, {\it Phys. Rev. Lett.} {\bf 83} (1999) 3370.


\bibitem{Dim} N. Arkani-Hamed, D. Dimopoulos, N. Kaloper and R. Sundrum, {it Phys. Lett. } {\bf B 480} (2000) 193; S. Kachru, M. Schulz and E. Silverstein, {\it Phys. Rev. }{\bf D62} (2000) 08 5003; S. Forste, Z. lalak, S. Lavignac and H. P. Nilles, {\it Phys. Lett. }{\bf B 481} (2000) 360.

\bibitem{ya} O. Aharony, S. S. Gubser, J. Maldacena, H. Ooguri and Y. Oz,
{\it Phys. Rept.} {\bf 323} (2000) 183.


\bibitem{ka} 
R. Altendorfer, J. Bagger and D. Nemeschansky, ``Supersymmetric Randall-Sundrum Scenario'', hep-th/0003117;
T. Gherghetta and A. Pomarol, {\it Nucl. Phys.} {\bf B 586} (2000) 141;
A. Falkowski, Z. Lalak and S. Pokorski, {\it Phys. Lett.}{\bf B491} (2000) 172.

\bibitem{pok} E. Bergshoeff, R. Kallosh and A. Van Proyen, {\it JHEP} 0010 (2000) 033; A. Falkowski, Z. Lalak and S. Pokorski,''Five-Dimensional Gauged Supergravities with Universal
Hypermultiplet and Warped Brane Worlds, hep-th/0009167.


\bibitem{luk} A. Lukas, B. A. Ovrut, K. S. Stelle, D. Waldram, {\it Phys. Rev. D}{\bf 59} (1999) 086001;
A. Lukas, B. A. Ovrut, K. S. Stelle, D. Waldram, {\it Nucl. Phys. }{\bf B552} (1999) 246;
K. Behrndt and S. Gukov, {\it Nucl. Phys. }{\bf B 580} (2000) 225.

\bibitem{ov} A. Lukas, B. A. Ovrut and D. Waldram, {\it Phys. Rev. D}{\bf 59} (1999) 106005.

\bibitem{ovr} A. Lukas, B. A. Ovrut and D. Waldram, ``Cosmology and Heterotic M-Theory in Five Dimensions'',
Lectures at the advanced school on Cosmology and Particle Physics, Peniscola, Spain, June 1998.



\bibitem{cosm} A. Lukas, B. A. Ovrut and D. Waldram, {\it Phys. Rev. D}{\bf 60} (1999) 086001;
M. Braendle, A. Lukas and B. A. Ovrut, {\it Phys. Rev.} {\bf D 63} (2001) 026003; H. A. Chamblin and H. S. Reall, {\it Nucl. Phys. }{\bf B562} (1999) 133; J. E. Lidsey,
{\it Class. Quant. Grav.} {\bf 17} (2000) L39.

\bibitem{senreview} A. Sen, ``Non-BPS states and Branes in String
  Theory'', APCTP winter school lectures, hep-th/9904207.



\bibitem{sentach} A. Sen, {\em Int.J.Mod.Phys.} {\bf A14} (1999) 4061.

\bibitem{sol} S.P. de Alwis and N. Irges, {\it Phys. Lett. } {\bf B492} (2000) 171.

\bibitem{green} B. R. Greene, K. Schalm and G. Shiu, ``Dynamical Topology Change in M Theory'',
hep-th/0010207.

\bibitem{brax} Ph. Brax and A. C. Davis, {\it Phys. Lett. }{\bf B497} (2001) 289.

\bibitem{pierre} P. Binetruy, J. Cline and C. Grojean, {\it Phys. Lett.} {\bf B 489} (2000) 403.
\bibitem{beck} K. Becker, M. Becker , D. R. Morrison, H. Ooguri, Y. Oz and Z. Yin, {\it Nucl. Phys. }{\bf B480} (1996) 225. 

\bibitem{gib} G. W. Gibbons and G. Papadopoulos, {\it Comm. Math. Phys.} {\bf 202} (1999) 593.

\bibitem{senK} J. Majumder and A. Sen, {\it JHEP} 0009 (2000) 047.

\bibitem{wit} E. Witten, {\it Nucl. Phys.} {\bf B 460} (1996) 541; B. Ovrut, T. Pantev and J. Park,
{\it JHEP} 0005 (2000) 045. 

\end{thebibliography}
\end{document}